Research article

**Open Access**

# Exploring nervous system transcriptomes during embryogenesis and metamorphosis in *Xenopus tropicalis* using EST analysis

Ana C Fierro*[1,2,5], Raphaël Thuret[1,2], Laurent Coen[3], Muriel Perron[1,2], Barbara A Demeneix[3], Maurice Wegnez[1,2], Gabor Gyapay[4], Jean Weissenbach[4], Patrick Wincker[4], André Mazabraud[1,2] and Nicolas Pollet*[1,2,5]

Address: [1]CNRS UMR 8080, F-91405 Orsay, France, [2]Univ Paris Sud, F-91405 Orsay, France, [3]CNRS UMR 5166, Evolution des Régulations Endocriniennes, USM 501, Département Régulations, Développement et Diversité Moléculaire, Muséum National d'Histoire Naturelle, 7 rue Cuvier, 75231 Paris Cedex 5, France, [4]Genoscope and CNRS UMR 8030, 2 rue Gaston Crémieux CP5706, 91057 Evry, France and [5]Programme d'Épigénomique, Univ Evry, Tour Évry 2, 10è étage, 523 Terrasses de l'Agora, 91034 Evry cedex, France

Email: Ana C Fierro* - carolina.fierro@gmail.com; Raphaël Thuret - raphael.thuret@u-psud.fr; Laurent Coen - coen@mnhn.fr; Muriel Perron - muriel.perron@u-psud.fr; Barbara A Demeneix - demeneix@mnhn.fr; Maurice Wegnez - maurice.wegnez@u-psud.fr; Gabor Gyapay - gabor@genoscope.cns.fr; Jean Weissenbach - jsbach@genoscope.cns.fr; Patrick Wincker - pwincker@genoscope.cns.fr; André Mazabraud - andre.mazabraud@u-psud.fr; Nicolas Pollet* - Nicolas.Pollet@u-psud.fr

* Corresponding authors

## Abstract

**Background:** The western African clawed frog *Xenopus tropicalis* is an anuran amphibian species now used as model in vertebrate comparative genomics. It provides the same advantages as *Xenopus laevis* but is diploid and has a smaller genome of 1.7 Gbp. Therefore *X. tropicalis* is more amenable to systematic transcriptome surveys. We initiated a large-scale partial cDNA sequencing project to provide a functional genomics resource on genes expressed in the nervous system during early embryogenesis and metamorphosis in *X. tropicalis*.

**Results:** A gene index was defined and analysed after the collection of over 48,785 high quality sequences. These partial cDNA sequences were obtained from an embryonic head and retina library (30,272 sequences) and from a metamorphic brain and spinal cord library (27,602 sequences). These ESTs are estimated to represent 9,693 transcripts derived from an estimated 6,000 genes. Comparison of these cDNA sequences with protein databases indicates that 46% contain their start codon. Further annotation included Gene Ontology functional classification, InterPro domain analysis, alternative splicing and non-coding RNA identification. Gene expression profiles were derived from EST counts and used to define transcripts specific to metamorphic stages of development. Moreover, these ESTs allowed identification of a set of 225 polymorphic microsatellites that can be used as genetic markers.

**Conclusion:** These cDNA sequences permit *in silico* cloning of numerous genes and will facilitate studies aimed at deciphering the roles of cognate genes expressed in the nervous system during neural development and metamorphosis. The genomic resources developed to study *X. tropicalis* biology will accelerate exploration of amphibian physiology and genetics. In particular, the model will facilitate analysis of key questions related to anuran embryogenesis and metamorphosis and its associated regulatory processes.

## Background

*Xenopus tropicalis* is now an anuran amphibian reference genome for vertebrate comparative genomics. It presents the same advantages as *Xenopus laevis* but has a smaller genome of 1.7 Gbp and a shorter generation time [1]. Moreover, while *X. laevis* is an allotetraploid derived from an allopolyploidization event, *X. tropicalis* is diploid [2,3]. Even though phylogenetic studies indicate that 30 to 50 MY evolution separate the two species [3,4], it has been shown that most methods and resources developed for *X. laevis* can be readily applied to *X. tropicalis* [5]. Thus, the genome of *X. tropicalis* was selected to explore amphibian genome characteristics by whole-genome shotgun sequencing [6].

Working on *X. laevis* constitutes a challenge when dealing with large-scale transcriptomics, such as microarrays experiments or systematic cDNA sequencing. This is because some *X. laevis* genes are present as diploids, while others form pairs of paralogs (also called "pseudoalleles") that have been conserved with various degrees of divergence, generally less than 10% [7]. On a genomic scale, recent data has led to the estimation of 12% as the minimal fraction of paralogous gene pairs kept after allotetraploidization [8]. However, this estimate is based on the application of strict and conservative criteria: less than 98% nucleotidic similarity and 93% mean similarity between paralogs. Therefore, it is likely that more than 12% of paralogs are indeed active genes in *X. laevis*. Moreover, such pairs of genes may have distinct expression patterns [7]. An estimated 14% of paralogs show distinct expression profiles based on EST counts [8]. Given these complications, it follows that the *X. tropicalis* genome is more amenable to systematic transcriptome surveys than that of *X. laevis*

Transcriptome analysis relies heavily on cDNA analysis. Collections of cDNA sequences have multiple uses for the molecular geneticist. They can be used to establish transcript catalogues [9-11] and to provide experimental evidence when building gene models from genomic sequence, particularly for 5' and 3' untranslated sequences [12]. Further, they can be used to provide global views on genome expression in a given cell type by the estimation of the abundance of the different mRNA species (through signatures as in [13]) and therefore can help decipher physiological roles played by a given gene product. Finally, partial cDNA sequences (ESTs) are used to identify full-length clones containing the entire open-reading frame for each transcript [14].

We initiated an EST program so as to provide a functional genomics resource for *X. tropicalis* containing sequences from the highest possible number of genes expressed in the nervous system. We report the construction of such a gene index and its assessment after the collection of 48.785 partial cDNA sequences. These ESTs are estimated to represent 6,000 genes that were annotated through sequence similarity searches, protein domain searches and Gene Ontology functional classification. Gene expression profiles were derived from EST counts and used to evidence transcripts differentially expressed at metamorphic stages of development. A set of polymorphic intragenic microsatellite markers was deduced from the analysis of ESTs derived from distinct strains of *X. tropicalis*. We expect that this resource will be valuable for further molecular genetics experiments.

## Results and discussion
### Construction of cDNA libraries and normalization

Two *X. tropicalis* cDNA libraries were constructed for this project. The first, designated xthr, was derived from dissected retinas and heads of young tadpoles (Nieuwkoop and Faber st. 25–35). About 500 retinas were dissected from stage 32 *X. tropicalis* embryos, a stage where differentiating retinal neurons are getting organized into layers. Because these retinas yielded only few polyA+ RNA, the library was enriched by the addition of mRNA from heads of embryos of the same developmental stage. The second library, designated xtbs, was made from central nervous systems of metamorphosing tadpoles. Brains and spinal cords were dissected from tadpoles between stage 58 and 64, the period covering the whole of *Xenopus* metamorphosis. To build the library, and with the aim of respecting the relative proportion of nervous tissue obtained at the different stages, samples for six animals were pooled for each stage between 58–61 and three animals for each stage between 62–64. All these tissues were combined and the mRNA extracted for preparation of the xtbs library. The SMART technology (Clontech) was used to enrich the representation of full-length cDNA clones (defined here as a copy of the transcript sequences between the 5' cap and a polyA tail).

To increase the information derived from EST projects, it is necessary to sample complex or normalised cDNA libraries with few overrepresented cDNA clones (observed individually with a frequency greater than 1%). To evaluate our libraries quality, samples of 1,989 cDNAs from xthr and 1,694 cDNA from xtbs were partially sequenced (see Methods) to obtain 4,120 ESTs. Next, a normalization step was performed to increase the diversity of sequence tags. We used a set of 53 oligonucleotides (35 mers) corresponding to highly represented clones (≥ 1%, see Methods) in hybridizations on high-density colony filters (See additional file 1). A total of 22,561 clones were scored as positives (20% in both libraries) with an estimated false positive level of 0 and 3%, and an estimated false negative level of 38 and 10% for xthr and xtbs libraries, respectively. The negatively scored clones were re-arrayed to further the project using both 5' and 3' sequencing. The further sequencing of cDNA clones provided 48,785 high-quality sequences derived from 27,806 clones after trimming 57,874 reads (including the 4,120 ESTs of the pre-normalisation step, Table 1, see Methods). Both 5' and 3' end sequences were read for 75% of the cDNA clones, therefore reducing the difficulties associated with EST clustering. Moreover, this strategy helps to determine the choice of a given cDNA clone for further experiments, whether it be full-length cDNA sequencing, overexpression studies or complementary RNA in vitro synthesis.

To determine if the normalization process was successful, the number of sequences containing each oligonucleotide probe was counted before and after normalization (Fig. 1). Before normalization, the 53 clusters from which the probes were derived accounted for 18% of the 4,120 ESTs. This fraction dropped to 1% after normalization, confirming the efficacy of the method. Of the 48 clusters corresponding to nuclear genes, 18 (37%) have 20 or more corresponding ESTs and 17 (35%) have 40 or more ESTs after normalization. We conclude that the abundance of ESTs after normalisation was sufficient in the majority of cases. Even though this strategy requires re-arraying, there is no bias due to insert length compared to normalization by re-association [15] and therefore constitutes a useful alternative.

*EST assembly*

We analyzed these sequences with PHRAP [16] to build contigs out of the overlapping and redundant sequences (Table 1). A total of 31,767 sequences were assembled into 8,756 contigs. These were further grouped by virtue of clone links into 6,547 unique groups (scaffolds). Taking into account the 2,982 singletons issued from 2,304 clones, a total of 9,693 transcripts sequences were identified. We compared our results to the global clustering of all *X. tropicalis* ESTs (including ours) by the UniGene pipeline and the DFCI Gene Index. In UniGene, our set of ESTs belong to 7,778 groups made of between 1 and 220 clones. Similarly, The DFCI *Xenopus tropicalis* Gene Index clustered these ESTs in 9,350 TCs and 1,160 singletons.

The majority of clusters (66%) contained three or less ESTs. Only 11 contigs were composed of more than 100 sequences (See Additional file 2) and the largest contig contained 159 sequences. Most of the corresponding gene products (23/50) are ribosomal proteins, the other being proteins involved in basic cellular processes (tubulin, elongation factor 1 alpha). Two noteworthy exceptions are myelin basic protein (contig8746) and metallothionein (contig8708), for which transcripts are found almost exclusively in the nervous sytem.

The sequence redundancy (number of ESTs/cluster) of xthr and xtbs libraries was compared to other *X. tropicalis* cDNA libraries represented in dbEST (See Additional file 3). A statistically significant difference at the 1% level of significance indicates that the complexity is higher for adult-type cDNA libraries, whether or not a size fractionation was performed. Amongst cDNA libraries prepared from embryonic or larval stages of development, the complexity of the xtbs library ranks first, while the complexity of the xthr library is close to the mean value.

Sequences were assembled into contigs of up to 3 kb in size (hsp90 transcript, Contig 8575, See Additional file 4), but the mean contig length of 745 bp indicated that most of them cover only parts of the cognate transcript sequence.

To assess the fraction of clones likely to be full-length, we estimated the number of sequences in our dataset that

**Table 1:** *Xenopus tropicalis* EST project statistics

|  | xthr | xtbs | xthr and xtbs |
|---|---|---|---|
| Number of sequences reads obtained | 30272 | 27602 | 57874 |
| Number of clone sequences obtained | 16548 | 14901 | 31449 |
|  |  |  |  |
| Number of valid sequences | 26440 | 22345 | 48785 |
| Number of clones with valid sequences | 15540 | 12266 | 27806 |
| Number of clones with 5' and 3' EST | 9354 | 11486 | 20840 |
| Number of clones with 5' EST only | 4831 | 0 | 4831 |
| Number of clones with 3' EST only | 1355 | 780 | 2135 |
| Average trimmed EST length | 522 | 546 | 534 |
|  |  |  |  |
| Number of contigs | 4327 | 4002 | 8756 |
| Number of contigs groups | 497 | 289 | 842 |
| Number of contigs grouped | 1210 | 649 | 2209 |
| Number of unique contigs | 3117 | 3353 | 6547 |
| Nulmber of clones in contigs | 9616 | 9268 | 15642 |
| Number of singletons | 7958 | 4262 | 17018 |
| Number of clones in singletons | 5924 | 2998 | 12164 |
| Number of putative transcripts | 9538 | 6640 | 19553 |
| Max. assembled sequence length | 3028 | 3144 | 3028 |
| Average assembled sequence length | 732 | 782 | 745 |
| Max. assembled sequence size | 147 | 144 | 159 |
| Average assembled sequence size | 6 | 5 | 5 |
| Number of contigs containing |  |  |  |
|  |  |  |  |
| 1 | 509 | 97 | 1152 |
| 2 | 1779 | 2155 | 3730 |
| 3 | 452 | 362 | 905 |
| 4–5 | 587 | 545 | 1159 |
| 6–10 | 505 | 476 | 952 |
| 11–20 | 283 | 207 | 479 |
| 21–30 | 91 | 90 | 174 |
| 31–50 | 80 | 41 | 118 |
| 50–100 | 31 | 26 | 76 |
| >100 | 10 | 3 | 11 |

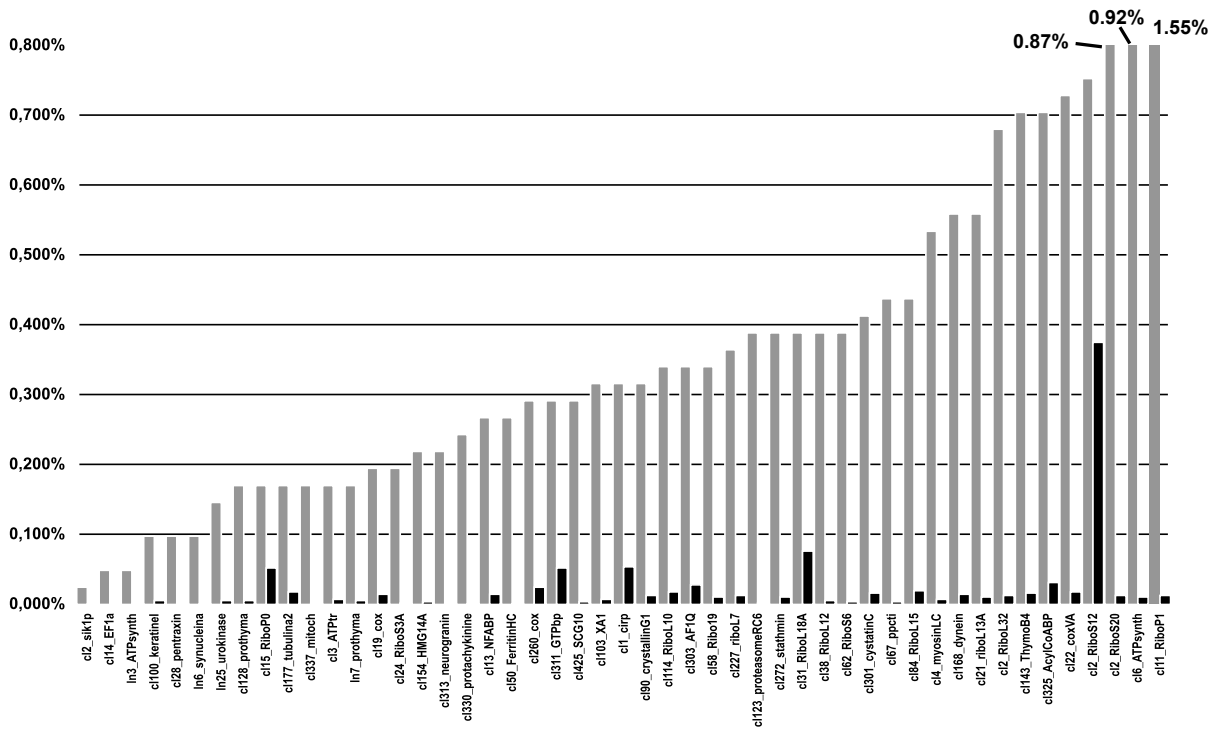

#### Figure 1
**Assessment of normalization effectiveness**. Histogram showing the percentage of sequence matching each oligonucleotide used in the procedure of normalization by hybridization. Bars represent percentages of positive clones calculated before (grey bars) and after normalization (black bars). Data before normalization were obtained after partial sequencing of 1,989 cDNAs from xthr and 1,694 cDNAs from xtbs. Note the relatively high abundance of cell-type specific transcript such as gamma crystallin (crystallinG1) or neurogranin (underlined on the figure).

extends over the 5' or 3' end of complete cDNA sequences (Figure 2; 1,945 entries from the *X. tropicalis Xenopus* Gene Collection [8], Xt-XGC and 2,963 entries from the Sanger Institute [17]). Using conservative criteria, at least one 5'EST was found to provide additional 5' upstream or 3' downstream sequence for 854 complete cDNAs (17.4% of the set). Using the same criteria but only on contig sequences, further sequence information was obtained on 355 complete cDNAs (7.2% of the set). Of these full-length cDNAs, 82 are completely matched by 122 contigs, and the latter are all longer. These results provide an indication of the added-value of this sequence resource in the framework of the delineation of gene structure, especially with respect to the determination of the transcription initiation site.

Another way to assess the fraction of clones likely to be full-length has been described by Gilchrist et al. [17]. Using this method on all *X. tropicalis* cDNA sequences (version Xt6 [18]) xthr and xtbs libraries we found to contain respectively 42% and 37% of full-length clones (MJ Gilchrist, personal communication). The mean fraction of full-length clones across all libraries is 18%. Therefore, we conclude that our normalization procedure did not impair the proportion of full-length clones compared to non-normalized libraries.

### Sequence annotations
In order to further analyse our dataset, we compared our contigs to ENSEMBL predicted transcripts. Altogether, 4,437 contigs (52%) and 1,423 singletons (48%) matched 4,083 transcripts from 3,703 ENSEMBL predicted genes (15%). The extent of the underclustering of our ESTs was estimated from these numbers and used to calculate that our whole EST set represents about 6,000 genes. We conclude that our cDNA sequence collection significantly improved annotation of the *X. tropicalis* genome sequence. Similarly, we compared our dataset to 2,402 *X. tropicalis* RefSeq mRNA sequences. We found that 2,230 contigs (26%) and 484 (16%) singletons matched 1,342 RefSeq entries (56%). These figures suggest that further extensive sequencing of putative full-length cDNA clones from our collection will be of great use in order to cover the entire *Xenopus* gene set.

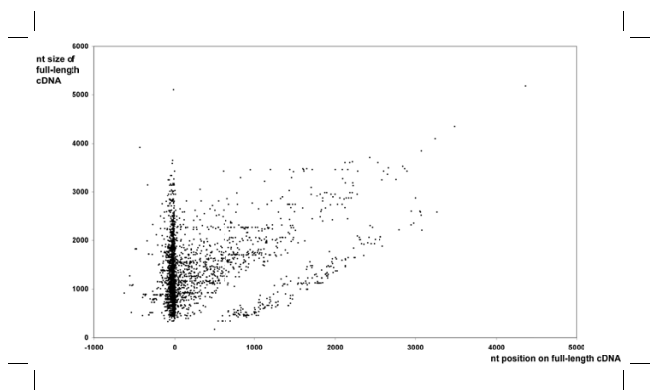

**Figure 2**
**Added-value of xthr and xtbs 5'ESTs**. 5' cDNA sequences were compared to 4,908 complete *X. tropicalis* cDNA sequences from XGC and Sanger Institute. When an EST matched unambiguously (>95% id over more than 50 nt on the same orientation) one of these cDNAs, the position of its first residue (X axis) was plotted as a function of the cDNA size (Y axis). Each dot represents the result of an alignment. A position of 0 on the x axis indicates identical 5' ends between the EST and cDNA. Negative values indicate that the EST extends further 5', positive values superior to the cDNA length indicate that the EST extends further 3', and positive values inferior to the cDNA size indicate the 5' EST position relative to the cDNA.

We next estimated the proportion of our cDNA sequences representing mRNA molecules produced by a splicing event and hence most likely to correspond to physiological products. We used "exonerate" to compute alignments between cDNA and genomic sequences. We retained only the alignments satisfying the thresholds of 95% identity and 90% coverage. Evidence of splicing was found for 5,025 contigs (65%) out of 7,718 contigs aligned to the genome. From the 2,693 contigs left, only 274 are significantly similar to a protein sequence and it is likely that the others represent 3' untranslated regions, often encoded by a single exon in vertebrates.

Next, two complimentary methods were used to find evidence for alternative splicing. Using genomic sequences (see Methods) we predicted 111 cases of alternative splicing, including conserved ones such as Clathrin light chain (contig7735 and contig8467, See Additional file 5). Using alignments on protein and genomic sequences we found 58 cases (such as *elrD* represented by contig7817, See Additional file 6).

Our set of transcript sequences was annotated using similarity searches in nucleotidic and protein databases and motif searches (Table 2). Of the 8,756 contigs, 62% have more than 70% nucleotidic similarity to previously described *X. laevis* regular entries, and may be considered as "known" *Xenopus* genes. Of these sequences, 4,426 had significant similarity to 2,803 protein sequences in Swiss-Prot database, and 5,506 to 3,571 cluster of the Uniref90 database. We identified 212 sequences corresponding to the *Xenopus* orthologs of human disease genes (See Additional file 7). Further molecular studies on these genes in *Xenopus* will be useful for understanding the physiopathology of these diseases.

Putative coding regions were identified using framesearch, and corresponding protein sequences were annotated using InterProScan, allowing for an automatic Gene Ontology Annotation (Table 3).

Several known genes specifically expressed in the eye were identified, including different crystallins (beta, gamma and mu isoforms), *vsx1* (visual system homeobox 1), *pax6* (paired-box protein 6), *rdgb* (retinal degeneration B homolog), *rgr* (RPE-retinal G protein-coupled receptor). Well-characterised central nervous system specific genes were identified as well, notably *elrC*, *mbp* (myelin basic protein), *plp* (myelin proteolipid protein 1). The corresponding cDNAs will provide useful differentiation markers for *X. tropicalis*.

A significant number of the contigs (37%) had no significant similarities to previously described genes, and may represent transcribed pseudogenes, non-coding RNA sequences and undescribed genes. Indeed, comparing our sequences to non-coding RNA sequences (microRNA from RFAM, or ncRNA from the H-INV datasets) we found 2 microRNA precursors (contig7127 and 7850 encoding mir-9-1 and mir-124a respectively) as well as E3 (Contig2965) and 5SN4 (Contig5668) snoRNAs. Contig7127 (452 nt) is derived from the assembly of 6 ESTs derived from 3 distinct cDNA clones of the xtbs library. The alignment of contig7127 sequence on *X. tropicalis* genome sequence reveals 100% identity and indicates that one splicing event is required. Thus, contig7127 represents a bona fide neural transcript of the mir-9-1 gene. Contig7850 (800 nt) is derived from the assembly of 10 ESTs derived from 6 cDNA clones (one from xtbs and 5 from xthr libraries). Four of these clones are identical and characterized by a 409 bp cDNA, while two are longer and have their 3' ends ESTs as singletons but mapping to the same scaffold region.

**Table 2: Results of sequence comparisons**

| | BLASTX | | | | BLASTN | | | | | | | | | |
|---|---|---|---|---|---|---|---|---|---|---|---|---|---|---|
| | SwissProt | | Uniref 90 | | X. laevis reg. entries | | X. laevis UniGene | | X. tropicalis UniGene | | X. tropicalis genome | | X. tropicalis cDNA | |
| | | | | | | | | | | | ND | | 2034 | | |
| All hits | 4426 | 51% | 5506 | 63% | 5447 | 62% | 7175 | 82% | 7865 | 90% | 8703 | 99% | 3877 | 44% | All hits |
| >= 90% | 1594 | 18% | 2753 | 31% | 744 | 8% | 864 | 10% | 7199 | 82% | 8371 | 96% | 3660 | 42% | >95% |
| 70 – 90% | 1705 | 19% | 1917 | 22% | 2963 | 34% | 3843 | 44% | 454 | 5% | 302 | 3% | 217 | 2% | 90 – 95% |
| < 70% | 1127 | 13% | 836 | 10% | 1740 | 20% | 2468 | 28% | 212 | 2% | 30 | 0% | 0 | 0% | < 90% |
| No similarity | 4330 | 49% | 3250 | 37% | 3309 | 38% | 1581 | 18% | 891 | 10% | 53 | 1% | 4879 | 56% | No similarity |

*Expression profiles*

Other collections of *X. tropicalis* ESTs are ongoing [8,17,19] using a variety of cDNA libraries made from adult tissues or embryos at different stages of development. Hence, we undertook an *in silico* analysis of gene expression profiles estimated from EST counts [20].

In a first analysis, we searched transcripts identified by ESTs derived predominantly from our cDNA libraries. We identified 99 and 238 cDNAs found prominently in the heads and retinas of tailbuds or brain and spinal cord of tadpole, respectively (See Additional file 8 and Additional file 9) and 25 clones found predominantly in both structures. These clones are likely to represent genes differentially expressed in the retina or the central or peripheral nervous system during metamorphosis. The study of these genes in *Xenopus* could well improve our knowledge on CNS development and function in vertebrates.

*Metamorphosis*

In a second analysis, we explored the metamorphosis transcriptomes using expression profiles derived from EST counts. It is known that amphibian metamorphosis brings about unique regulations triggered by thyroid hormones during late vertebrate development, but relatively few genes are characterised as playing regulatory roles in this process [21]. We extracted 4,187 UniGene clusters containing at least one EST from the xtbs cDNA library. Similarly, we fetched data from 592 UniGene clusters containing at least one and up to 132 ESTs from another library derived from a metamorphic stage of development. Combining both sets gives 4,779 UniGene clusters (13% of all clusters). To generate a useful expression matrix an initial filtration step was performed whereby clusters composed of less than 10 ESTs were removed leading to a set of 3,422 UniGene clusters. We used the GT test [22] to rank profiles in three categories: strong (64 clusters with GT > 0.66), medium (803 clusters with 0.33 < GT < 0.66) or weak (2555 cluster with GT < 0.33) differential expression. Because UniGene is prone to overclustering we focused our analysis on the 64 clusters corresponding to genes with a strong differential expression and analysed their expression profiles using hierarchical clustering (Figure 3). Twelve characteristic expression profiles are observed, corresponding to peaks of expression that are tissue (brain, intestine, kidney, heart, lung, skeletal muscle, skin) or stage-specific (egg, tailbud, tadpole, metamorphosis). The corresponding genes are potential differentiation markers that can be useful in developmental studies and can easily be checked by in situ hybridization on embryos. Only one transcript tagged by ESTs derived solely from a metamorphic stage was identified. This transcript codes for preprocaerulein type-4; it is characterized by 40 ESTs derived from 24 cDNA clones issued from 6 libraries made from stage 62 and 64 tadpoles, i.e. representing late metamorphosis stages. Caerulein is a peptide found predominantly in skin secretions. It belongs to the gastrin/cholecystokinin family of neuropeptide, and may play a role as an antimicrobial molecule [23]. This finding is discussed later.

Since there are currently ten times more ESTs in cDNA libraries derived from metamorphic stages of development in *X. laevis* than in *X. tropicalis*, we did a similar survey of the expression profiles of transcripts in *X. laevis*.

We extracted 6,297 UniGene clusters (24% of all clusters) containing at least one and up to 710 ESTs from at least one cDNA library prepared from metamorphic tadpoles. This corresponds to 24,262 ESTs made from four cDNA libraries: limb, tail, intestine and tadpole (NF stage 62). The level of expression of each transcript was estimated by counting ESTs providing a corresponding UniGene cluster. The 26 clusters containing the highest number of ESTs, and hence corresponding to the most highly expressed genes during metamorphosis, are listed in table 4. We expected to find either ubiquitously or differentially expressed categories among these highly abundant transcripts at metamorphosis stages. Indeed, 16 of these 25 UniGene clusters are found as characterized by a restricted expression in the tail, limb, or heart (table 4). Interestingly, it is known that 11 are expressed in the muscle cells that compose most of the tail and limb. One remarkable case is a gene coding for a protein involved in freeze-toler-ance found predominantly in metamorphic limbs, but expressed in a variety of other tissues both embryonic (starting at gastrula stage) and adult (nearly all adult tissues sampled with the exception of ovary, testis and lung). This can be an artefact due to the handling of tissues at the time of RNA extraction. Alternatively, this may reflect the induction by stress-related hormones (glucocorticoids) during metamorphosis.

**Table 3: GO Molecular function classification.**

| Gene ontology term | N |
| --- | --- |
| All molecular function terms | 1775 |
| ->antioxidant activity | 14 |
| ->binding | 623 |
| -->calcium ion binding | 89 |
| -->carbohydrate binding | 7 |
| -->lipid binding | 10 |
| -->nucleic acid binding | 350 |
| --->DNA binding | 135 |
| ---->chromatin binding | 0 |
| ---->transcription factor activity | 31 |
| --->RNA binding | 40 |
| --->translation factor activity, nucleic acid binding | 58 |
| -->nucleotide binding | 51 |
| -->oxygen binding | 0 |
| -->protein binding | 52 |
| --->cytoskeletal protein binding | 25 |
| ---->actin binding | 23 |
| -->receptor binding | 28 |
| ->catalytic activity | 455 |
| -->hydrolase activity | 110 |
| --->nuclease activity | 9 |
| --->peptidase activity | 66 |
| --->phosphoprotein phosphatase activity | 6 |
| -->kinase activity | 56 |
| --->protein kinase activity | 33 |
| -->transferase activity | 112 |
| ->chaperone regulator activity | 0 |
| ->enzyme regulator activity | 37 |
| ->motor activity | 15 |
| ->nutrient reservoir activity | 0 |
| ->signal transducer activity | 48 |
| -->receptor activity | 13 |
| -->receptor binding | 28 |
| ->structural molecule activity | 398 |
| ->transcription regulator activity | 40 |
| ->translation regulator activity | 58 |
| ->transporter activity | 120 |
| -->electron transporter activity | 39 |
| -->ion channel activity | 11 |
| -->neurotransmitter transporter activity | 3 |
| ->triplet codon-amino acid adaptor activity | 0 |

We then carried out an in silico reconstruction of the transcriptional profile of *X. laevis* metamorphic genes using the IDEG suite of statistical tests. We removed clusters composed of less than 10 ESTs, leading to a set of 3,599 UniGene clusters. The GT test ranked profiles in three categories: strong (167 clusters with GT > 0.66) medium (1,300 clusters with 0.33 < GT < 0.66) or weak (2,132 cluster with GT < 0.33) differential expression.

From the 167 clusters corresponding to genes with a strong differential expression, only 30 are composed of at least 2 ESTs derived solely from metamorphic or adult tissues, four of which bear no similarity to known proteins. We analysed the expression profiles of these metamorphic genes using unsupervised hierarchical clustering. The resulting clusters could be interpreted along the predominant expression domains (See Additional file 10). Three clusters (metamorphic, tadpole and limb) correspond to larval stages of development that are made of eight, three and four genes, respectively (Fig. 4). These genes are promising candidates, potentially playing important roles during this late developmental event. Below, we describe briefly what is known about each of these genes.

A larval beta chain of globin is among the metamorphic cluster together with an alpha chain, an indication of the relevance of our analysis. The comparison of Xl.56714 EST sequences with known proteins shows that they resemble cell surface receptors of the SLAM (Signalling Lymphocytic Activation Molecule) family. The SLAM receptors regulate immune cell activation. Indeed, it is known that immune system remodelling is a major event of metamorphosis [24]. The gene corresponding to Xl.56714 is expressed in metamorphic tadpoles (including tail and intestine) as well as in the adult kidney. However we could not detect significant similarities to any known gene sequences or proteins. Alpha-1 antichymotrypsin (a plasma protease inhibitor) is highly expressed during metamorphosis and found in adult liver. This correlates with the associated stress condition that occurs during tadpole transformations. The gene encoding alpha-2-HS-glycoprotein (also named fetuin) steadily increases in expression from tailbud stage up to metamorphosis. This gene product is secreted in plasma and plays a physiological role during mammalian fetal development, especially in mineralization and growth. A known *Xenopus* gene encoding a small peptide named PYLa is found exclusively in a cDNA library prepared from stage 62 tadpoles. As for the preprocaerulein transcripts in *X. tropicalis*, the PYLa transcripts are abundant in metamorphic stages, with ESTs found in limbs and whole tadpoles. Both caerulein and PYLa peptides may be secreted from skin glands and exert antimicrobial activities. This finding corroborates a previous report on caerulein expression [25]. Remarkably, skin glands are known to express a cocktail of signalling peptides, including neuropeptides such as xenopsin, thyrotropin-releasing hormone and PGLa. Whether these peptides play specific roles in the context of metamorphosis is unknown. The cluster Xl.24674 corresponds to a gene resembling uromodulin.

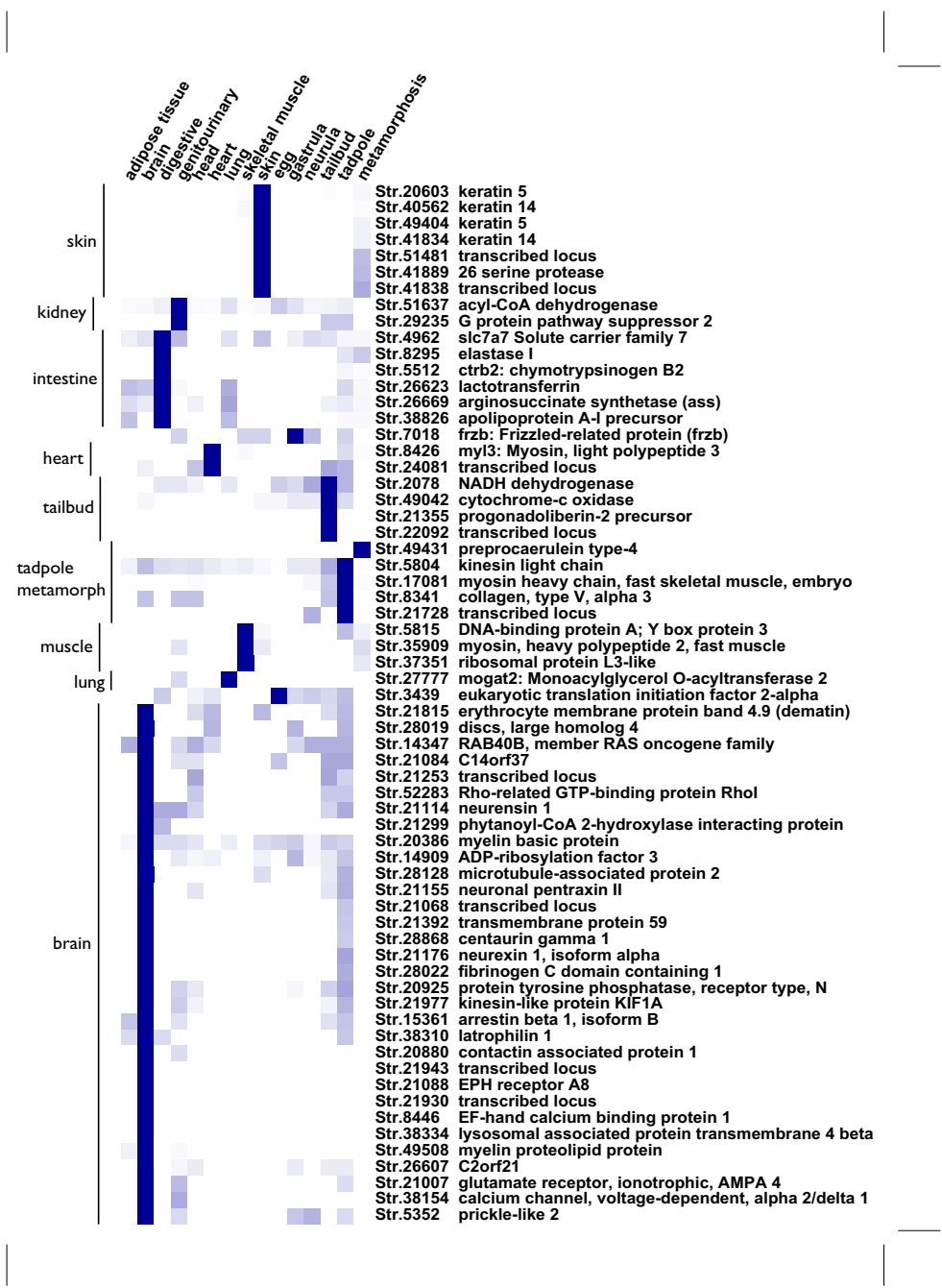

#### Figure 3
**Digital expression profiles of *X. tropicalis* transcripts differentially expressed at metamorphosis**. Each line gives the expression profile of a given transcript represented by a UniGene cluster. The expression is deduced from counting the occurence of ESTs derived from a given cDNA library. The level of expression is colour coded in blue shades, dark blue means evidence for high levels of transcripts and white means no evidence for expression. On the left, clusters of expression profiles are delineated by a vertical bar labelled with the associated characteristic domain of expression. On the right, the cluster name and its annotation (i.e. the corresponding gene product description as deduced from sequence similarity analysis) are given. Each column corresponds to a category of tissue or stage of development: 8 adult tissues and 6 stages of development. Note that a given category may correspond to several cDNA libraries. Here, only clusters for which evidence of differential expression were used to build the matrix of expression. This matrix was analysed by hierarchical clustering on the expression profile dimension using CLUSTER 3.0 as described in the methods section.

Table 4: Highly expressed metamorphic transcripts.

| UniGene ID | Met ESTs | Total ESTs | cDNA source | | Note | Restricted expression | Description |
|---|---|---|---|---|---|---|---|
| Xl.24656_a | 710 | 1405 | b;h;l;li;ot;sk;t;wb | R | ubiquitous | tail | *actin alpha 1 skeletal muscle* |
| Xl.1115__b | 248 | 653 | et;h;he;l;ot;sk;t;wb | R | tail, limb | tail | *actin alpha 1 skeletal muscle* |
| Xl.17432 | 560 | 4290 | b;et;ey;fb;h;he;k;l;li;lu;ot;ov;s;sk;t;te;th;wb | U | ubiquitous | - | eukaryotic translation factor 1 alpha, somatic form |
| Xl.5860 | 397 | 1008 | b;he;l;ot;s;sk;t;te;wb | U | ubiquitous | - | creatine kinase, muscle |
| Xl.11405 | 280 | 751 | b;et;l;t;wb | R | tail, limb, tailbud | tail | Calcium ATPase at 60A, cardiac muscle |
| Xl.24815 | 238 | 510 | b;l;t;th;wb | R | limb, tadpole | limb | myosin, heavy polypeptide 13, skeletal muscle |
| Xl.47042 | 227 | 468 | b;et;ey;fb;he;k;l;li;s;sk;t;th;wb | R | limb | metamorphosis | collagen, type I, alpha 1 |
| Xl.25492 | 217 | 833 | l;ot;sk;wb | R | limb, tadpole | limb metamorphosis | oncomodulin |
| Xl.7551 | 211 | 1333 | b;et;ey;fb;h;he;k;l;li;lu;ot;ov;s;sk;t;te;th;wb | U | ubiquitous | - | eukaryotic translation elongation factor 2 |
| Xl.28832 | 205 | 474 | et;h;h;l;t;wb | R | tail, limb | tail metamorphosis | actin alpha skeletal muscle |
| Xl.4138__a | 198 | 1376 | b;et;ey;fb;h;k;l;li;lu;ot;ov;s;sk;t;te;th;wb | U | ubiquitous | - | *actin cytoplasmic 1* |
| Xl.29221_b | 98 | 644 | b;et;ey;fb;h;he;k;l;li;lu;ot;ov;s;sk;t;te;th;wb | U | thymus | - | *actin cytoplasmic 1* |
| Xl.5146 | 168 | 360 | b;et;ey;fb;he;k;l;li;s;t;th;wb | R | limb | limb metamorphosis | freeze tolerance-associated protein FR47 |
| Xl.8842 | 164 | 1033 | b;et;ey;fb;h;he;k;l;li;lu;ot;ov;s;sk;t;te;th;wb | N | limb, thymus, spleen_PHA | - | ribosomal protein L3 |
| Xl.1032 | 158 | 243 | b;l;ot;sk;t;wb | R | tadpole | limb metamorphosis | fast skeletal troponin C beta |
| Xl.1055 | 151 | 288 | b;et;h;l;ot;sk;t;th;wb | R | limb, tailbud, tadpole | limb metamorphosis | myosin light chain 1, fast skeletal muscle isoform |
| Xl.24699 | 134 | 426 | b;ey;he;l;lu;t;th;wb | R | limb, heart | heart metamorphosis | actin alpha cardiac |
| Xl.7875 | 120 | 975 | b;et;ey;fb;h;he;k;l;lu;ot;ov;s;sk;t;te;th;wb | D | ubiquitous | - | solute carrier family 25 (mitochondrial carrier; adenine nucleotide translocator), member 5 |
| Xl.1728 | 120 | 884 | b;et;ey;fb;h;he;k;l;li;lu;ot;ov;s;sk;t;te;th;wb | N | limb, tadpole, olfactory epith. | - | acidic ribosomal protein P0 |
| Xl.8672 | 116 | 756 | b;et;ey;fb;h;he;k;l;li;lu;ot;ov;s;sk;t;te;th;wb | N | limb | - | ribosomal protein L4 |
| Xl.49126 | 115 | 272 | b;ey;fb;l;ot;s;sk;t;te;wb | D | ?ubiquitous | limb metamorphosis | procollagen, type I, alpha 2 |
| Xl.995 | 113 | 440 | b;et;ey;fb;h;he;k;l;lu;ot;ov;s;sk;t;te;th;wb | N | limb | - | glyceraldehyde-3-phosphate dehydrogenase |
| Xl.3463 | 109 | 554 | b;et;ey;fb;h;he;k;l;li;lu;ot;ov;s;sk;t;te;th;wb | R | limb | metamorphosis | guanine nucleotide binding protein, beta 2, related sequence 1 |
| Xl.118 | 102 | 274 | b;et;fb;h;k;l;lu;ot;s;sk;t;wb | R | limb | metamorphosis | tropomyosin 1 alpha chain |
| Xl.1464 | 100 | 537 | fb;l;li;ot;s;sk;t;wb | R | tail, tadpole | tail | myosin, heavy polypeptide 4, skeletal muscle |
| Xl.395 | 94 | 479 | li;lu;th;wb | R | liver | liver | serum albumin 74 kDa |
| Xl.938 | 94 | 221 | b;ey;fb;l;ot;sk;t;te;wb | R | limb | limb metamorphosis | tropomyosin beta chain, skeletal muscle |
| Xl.23399 | 94 | 156 | b;l;ot;sk;t;wb | D | ?ubiquitous | tail metamorphosis | aspartyl beta-hydroxylase |

b: brain;et: embryonic tissue;ey: eye;fb: fat body;h: head;he: heart;k: kidney;l: limb;li: liver;lu: lung;ot: other;ov: ovary;s: spleen;sk:skin; t: tail;te: testis;th: thymus;wb: whole body. Allogenes are written in italics. U: ubiquitous; D: no differential; N: no restricted expression; R: restricted expression.

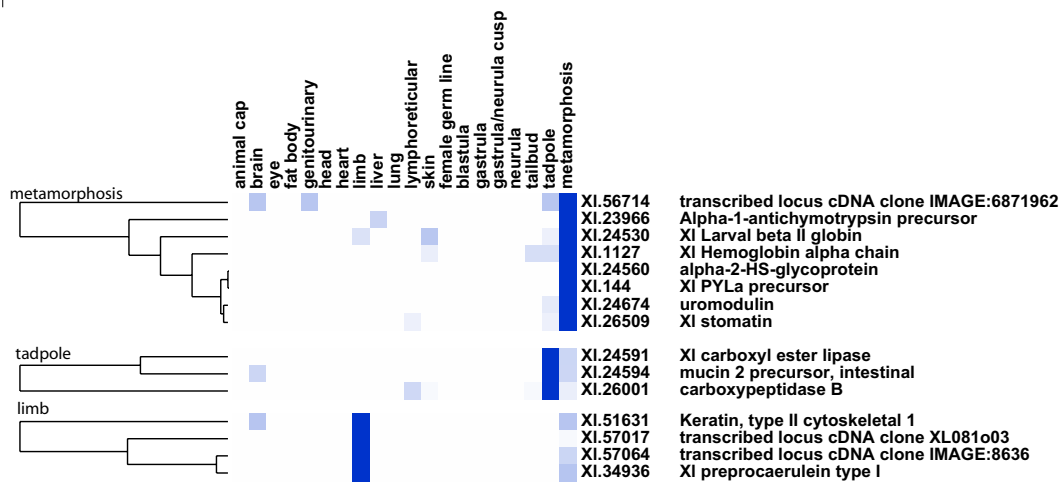

**Figure 4**
**Digital expression profiles of *X. laevis* transcripts differentially expressed at metamorphosis**. Using the same representation as in Fig. 3 three clusters are depicted that are associated with differential expression at metamorphosis (top cluster), tadpole stage (middle) and in the forming limb (down).

Corresponding transcripts are found in metamorphic intestine and whole tadpole. In mammals, uromodulin is excreted in urine and plays a role in the cellular defense response. A gene encoding a stomatin homolog is highly expressed in intestine during metamorphosis. Stomatin is a membrane protein regulating cation exchange and cytoskeletal attachment. Among the genes represented in the metamorphic limb cluster are a keratin and two clusters (Xl.57017 and Xl.57064) annotated as lacking significant similarities to known proteins. In the tadpole cluster, a carboxyl ester lipase is found expressed in tadpole and in metamorphic intestine. Mucin 2 is another gut protein highly expressed in tadpole, as well as carboxypeptidase.

Taken together, these expression profiles, based on EST counts, reveals certain genes that are up-regulated during metamorphosis, possible targets of the thyroid hormones signalling pathway.

*Polymorphisms*
The cDNA sequences we produced are derived from the Adiopodoume strain of *X. tropicalis*, originating from the Ivory Coast. The genomic and most other cDNA sequencing efforts are made on the N strain from Nigeria or a distinct IC strain from Ivory Coast. We therefore looked for polymorphisms that could be used in genetic mapping experiments or to discriminate with mutations obtained from ENU mutagenesis. We identified 8 SNPs derived from mitochondrial genes, three of which (snp1A, snp2A, snp6G) are specific to the Adiopodoume strain (See Additional file 11). The presence of shared alleles for 5 SNPs indicates the close relationship with the N strain as already reported by Evans et al. 2004. We searched for novel polymorphism markers made of di, tri, tetra and pentanucleotide sequence repeats present in our EST collection. We found from two to ten alleles in 225 markers derived from 212 contigs/ESTs clusters. A subset of 107 markers are potential highly informative since two or more alleles are observed at high frequencies (See Additional file 12). The dinucleotide repeat AT and TA are the most common, accounting for 137 markers. These intragenic markers should be useful once placed on a genetic linkage map.

This dataset will provide an invaluable tool for exon definition when the *X. tropicalis* genome sequence is finally determined. The results presented here are available through a database on our web site [26]. Users can carry out BLAST and other searches based on GO classification, InterproScan results, and expression information. The cDNA sequences have been deposited with Genbank/EMBL/DDBJ (accession numbers CN072222 – CN121006) and clones are available upon request.

**Conclusion**
Large-scale cDNA sequencing has provided invaluable resources to decipher vertebrate genome structure and function. Recent studies on cDNA sequencing with deep coverage provide fundamental knowledge on the complexity of transcriptomes in mammals [27]. Here, we provide information on the transcript sequence and expression of an estimated 6,000 genes in *X. tropicalis*. A web resource [26] is available with associated annotations. The genetic resources stemming from the cDNA sequencing project described here can be used in diverse research projects, including vertebrate comparative genomics, studies on evolution and development, cell biology and developmental genetics. More specifically, retinogenesis and remodelling of the central nervous sytem during metamorphosis will benefit from this cDNA resource.

We are currently undertaking full cDNA insert sequencing for a set of non-redundant clones, as well as characterizing their expression using a whole-mount in situ hybridization screen [28,29]. The genomic resources developed to study *X. tropicalis* biology are crucial to explore amphibian physiology and genetics, this model system providing excellent characteristics for addressing key questions related to anuran metamorphosis and its associated regulatory processes.

## Methods
### Embryo and tissue dissection
Embryos of *Xenopus tropicalis* Adiopodoumé strain were obtained from parents issued of the Geneva collection [30].

From each of the two libraries made, 58,368 clones were picked, arrayed in microtiter plates and gridded on high-density nylon filters. A sample of 1,989 cDNA clones from the xthr library and 1,694 clones from xtbs were partially sequenced to obtain 4,120 ESTs. This step provided a quality assessment of the two libraries, showing the absence of clones of bacterial origin and few ribosomal (0.45% in xthr, 0% in xtbs) and mitochondrial contaminants (1.2 and 3.6% in xthr and xtbs, respectively). This procedure provided information on overrepresented clones, which were then removed before further sequencing of up to 30,000 clones. These ESTs were grouped into 1,985 clusters.

### Normalization of cDNA libraries
Two pools of 25 and 28 oligonucleotides probes of 35 nt in length were labelled using Terminal Transferase (New England Biolabs) and $P^{33}$-dATP (Amersham). Labelled oligonucleotides were hybridised on two high-density filters representing the xthr and xtbs libraries as in [31]. Positive clones were identified using X-Digitize software on images acquired using a phosphorimager.

### High-throughput sequencing, assembling
The reactions were performed with a Big-Dye terminator cycle sequencing kit and analyzed by ABI-3700 and ABI-3730. Sequences were base-called using PHRED, then trimmed using LUCY and custom perl scripts. Sequences less than 100 bp were discarded, as well as those identified to be derived from ribosomal and mitochondrial RNAs. PHRAP was used to assemble the sequences taking into account quality scores, further clustering was obtained by scaffolding using mate-pairs informations. We retained scaffolds only if two clone links were available (excepting singletons) and if the orientation of the reads was consistent.

### Annotation
Repetitive sequences were masked using CENSOR. Contigs and singletons were used as queries in BLASTX and BLASTN searches of Swissprot, Uniprot, Unigene (rel 70) and *Xenopus tropicalis* JGI assembly v4.1 databases on INFOBIOGEN server. The february 2005 release of ENSEMBL was used. Framefinder was used to identify coding sequences, and protein domains were searched using INTERPROSCAN. The results of sequence assembly, scaffolding and annotation were loaded on a custom-made mySQL database. Web interface was developed using PHP scripts.

### Assessment of the fraction of clones likely to be full-length
5'ESTs or contigs were compared to full-length cDNAs using BLASTN. Alignments longer than 50 nt and exhibiting more than 95% identities were selected. Overlap between query and subject sequences was scored only when the alignment encompassed up to the last nucleotide at the 5' or 3' end of the sequences.

### Alternative splicing
Alignments between cDNA and genomic sequences (masked) were computed using exonerate. Evidence for alternative splicing was found when two contigs were aligned to the same genomic region with at least 95% mean overlap but with a different number of exons.

Alignments between cDNA and UNIREF protein sequences were computed using BLASTX. Alignments characterized by a gap (at least 10 aa) introduced in the contig sequence were retrieved. Alternative splicing was considered present if the contig sequence including the gap could be aligned to the genomic sequence.

### Expression profiles
EST counts were downloaded from UniGene (release 70 for *X. laevis* and 32 for *X. tropicalis*). Relevant profiles were extracted using custom perl scripts. GT test was run using the IDEG6 software [22]. Single hierarchical clustering was performed using Cluster 3.0 software [32]. We used absolute correlation similarity metrics followed by complete clustering on mean centered gene expression profiles. Results were visualised using TreeView.

### Polymorphisms
Mitochondrial SNPs were identified on a collection of mitochondrial ESTs downloaded from dbEST, JGI [33] and from our own set. These ESTs were assembled using CAP3 and analyzed using visualization software [34]. Microsatellites were identified using custom perl scripts.

## Authors' contributions
RT, LC and MP carried out laboratory and data analysis. AF wrote and ran the EST processing pipeline, including EST assembly and annotation, and is responsible for the web-available database. G.G., J.W. and P.W. managed and conducted the sequencing experiments. BD, MW and AM participated in the coordination of the study. NP participated in the conception and design of the study, carried out laboratory and data analysis and drafted the manuscript.

## Additional material

### Additional File 1
*Oligonucleotides used for normalization*. The table provided lists the oligonucleotide identifier, corresponding gene, number of corresponding ESTs, oligonucleotide sequence and Tm.
Click here for file
[http://www.biomedcentral.com/content/supplementary/1471-2164-8-118-S1.xls]

### Additional File 2
*Top 50 of contigs according to the number of constituent ESTs*. The table provided lists the Contig identifier, the number of constituent sequence reads, the contig length and the description of the best Swissprot hit identified by blastx.
Click here for file
[http://www.biomedcentral.com/content/supplementary/1471-2164-8-118-S2.xls]

### Additional File 3
*Analysis of* X. tropicalis *cDNA libraries complexity*. The data provided represent the description and complexity of X. tropicalis cDNA libraries sampled by more than 20,000 ESTs.
Click here for file
[http://www.biomedcentral.com/content/supplementary/1471-2164-8-118-S3.xls]

### Additional File 4
*Top 50 of contigs according to their size*. The table provided lists the Contig identifier, the contig length and the description of the best Swissprot hit identified by blastx.
Click here for file
[http://www.biomedcentral.com/content/supplementary/1471-2164-8-118-S4.xls]

### Additional File 5
*Alternative splicing detected by a different number of exons in contigs aligned in the same genomic region*. The table provided lists the Contig identifier, the number of constituent exons, the description of the best protein hit identified by blastx. Each line describes one alternative splicing case.
Click here for file
[http://www.biomedcentral.com/content/supplementary/1471-2164-8-118-S5.xls]

### Additional File 6
*Alternative splicing detected by a gap in the alignment against UniRef100*. The table provided lists the Contig identifier, the identifier of the protein evidencing an alternative splicing event, and the identifier of the protein showing the highest similarity by blastx. Each line describes one alternative splicing case.
Click here for file
[http://www.biomedcentral.com/content/supplementary/1471-2164-8-118-S6.xls]

### Additional File 7
Xenopus tropicalis *genes related to human disease genes*. The table provided lists the human disease trait, corresponding human gene name, chromosomal location, OMIM identifier and the corresponding Xenopus tropicalis Contig identifier.
Click here for file
[http://www.biomedcentral.com/content/supplementary/1471-2164-8-118-S7.xls]

### Additional File 8
*cDNA clones found specifically in library xthr*. The table provided lists the Contig identifier and the description of the best protein hit identified by blastx.
Click here for file
[http://www.biomedcentral.com/content/supplementary/1471-2164-8-118-S8.xls]

### Additional File 9
*cDNA clones found specifically in library xtbs*. The table provided lists the Contig identifier and the description of the best protein hit identified by blastx.
Click here for file
[http://www.biomedcentral.com/content/supplementary/1471-2164-8-118-S9.xls]

### Additional File 10
*Digital expression profiles of* X. laevis *transcripts*. Using the same formalism as in Fig. 3, all X. laevis Unigene clusters associated with differential expression at metamorphosis are depicted.
Click here for file
[http://www.biomedcentral.com/content/supplementary/1471-2164-8-118-S10.eps]

### Additional File 11
*Mitochondrial SNPs*. The table lists the occurence of given alleles of mitochondrial SNPs in the adiopodoume and N strains, in association with the corresponding mitochondrial gene.
Click here for file
[http://www.biomedcentral.com/content/supplementary/1471-2164-8-118-S11.xls]

### Additional File 12
*Highly informative intragenic microsatellite markers. The table lists allelic data for a set of intragenic microsatellite markers, including the Contig ID, corresponding UniGene cluster ID, number of alleles, type of microsatellite. In bold case are figured contig/ESTs/UG clusters for which at least two alleles have a frequency higher than the mean (calculated as the total number of ESTs divided by the number of alleles) or higher than 33%. Alleles number and frequency are shown in bold if the frequency is higher than the mean or higher than 33%. A * is indicative of more than one repeat polymorphism observed for that cluster.*
Click here for file
[http://www.biomedcentral.com/content/supplementary/1471-2164-8-118-S12.xls]

### Acknowledgements
We thank L. Du Pasquier for the gift of *X. tropicalis* animals and his continuous support. This research was funded by grants from l'Association pour la Recherche contre le Cancer, le Centre National de la Recherche Scientifique, le Ministère de l'Education, de la Recherche (French Xenopus Stock Center) et de la Technologie and the University of Paris Sud.

### References
1. Amaya E, Offield MF, Grainger RM: **Frog genetics: Xenopus tropicalis jumps into the future.** *Trends Genet* 1998, **14(7):**253-255.
2. Bisbee CA, Baker MA, Wilson AC, Haji-Azimi I, Fischberg M: **Albumin phylogeny for clawed frogs (Xenopus).** *Science* 1977, **195(4280):**785-787.
3. Evans BJ, Kelley DB, Tinsley RC, Melnick DJ, Cannatella DC: **A mitochondrial DNA phylogeny of African clawed frogs: phylogeography and implications for polyploid evolution.** *Mol Phylogenet Evol* 2004, **33(1):**197-213.
4. Evans BJ, Kelley DB, Melnick DJ, Cannatella DC: **Evolution of RAG-1 in polyploid clawed frogs.** *Mol Biol Evol* 2005, **22(5):**1193-1207.
5. Khokha MK, Chung C, Bustamante EL, Gaw LW, Trott KA, J. Y, Lim N, Lin JC, Taverner N, Amaya E, Papalopulu N, Smith JC, Zorn AM, Harland RM, Grammer TC: **Techniques and probes for the study of Xenopus tropicalis development.** *Dev Dyn* 2002, **225:**499-510.
6. Richardson P, Chapman J: **The Xenopus tropicalis genome project.** *Current Genomics* 2003, **4:**645-652.
7. Graf JD, Kobel HR: **Genetics of Xenopus laevis.** *Methods Cell Biol* 1991, **36:**19-34.
8. Morin RD, Chang E, Petrescu A, Liao N, Griffith M, Chow W, Kirkpatrick R, Butterfield YS, Young AC, Stott J, Barber S, Babakaiff R, Dickson MC, Matsuo C, Wong D, Yang GS, Smailus DE, Wetherby KD, Kwong PN, Grimwood J, Brinkley CP 3rd, Brown-John M, Reddix-Dugue ND, Mayo M, Schmutz J, Beland J, Park M, Gibson S, Olson T, Bouffard GG, Tsai M, Featherstone R, Chand S, Siddiqui AS, Jang W, Lee E, Klein SL, Blakesley RW, Zeeberg BR, Narasimhan S, Weinstein JN, Pennacchio CP, Myers RM, Green ED, Wagner L, Gerhard DS, Marra MA, Jones SJ, Holt RA: **Sequencing and analysis of 10,967 full-length cDNA clones from Xenopus laevis and Xenopus tropicalis reveals post-tetraploidization transcriptome remodeling.** *Genome Res* 2006, **16(6):**796-803.
9. Adams MD, Kerlavage AR, Fleischmann RD, Fuldner RA, Bult CJ, Lee NH, Kirkness EF, Weinstock KG, Gocayne JD, White O, *et al.*: **Initial assessment of human gene diversity and expression patterns based upon 83 million nucleotides of cDNA sequence.** *Nature* 1995, **377(6547 Suppl):**3-174.
10. Houlgatte R, Mariage-Samson R, Duprat S, Tessier A, Bentolila S, Lamy B, Auffray C: **The Genexpress Index: a resource for gene discovery and the genic map of the human genome.** *Genome Res* 1995, **5(3):**272-304.
11. Marra M, Hillier L, Kucaba T, Allen M, Barstead R, Beck C, Blistain A, Bonaldo M, Bowers Y, Bowles L, Cardenas M, Chamberlain A, Chappell J, Clifton S, Favello A, Geisel S, Gibbons M, Harvey N, Hill F, Jackson Y, Kohn S, Lennon G, Mardis E, Martin J, Mila L, McCann R, Morales R, Pape D, Person B, Prange C, Ritter E, Soares M, Schurk R, Shin T, Steptoe M, Swaller T, Theising B, Underwood K, Wylie T, Yount T, Wilson R, Waterston R: **An encyclopedia of mouse genes.** *Nat Genet* 1999, **21(2):**191-194.
12. Wei C, Brent MR: **Using ESTs to improve the accuracy of de novo gene prediction.** *BMC Bioinformatics* 2006, **7:**327.
13. Okubo K, Hori N, Matoba R, Niiyama T, Fukushima A, Kojima Y, Matsubara K: **Large scale cDNA sequencing for analysis of quantitative and qualitative aspects of gene expression.** *Nat Genet* 1992, **2(3):**173-179.
14. Gomez SM, Eiglmeier K, Segurens B, Dehoux P, Couloux A, Scarpelli C, Wincker P, Weissenbach J, Brey PT, Roth CW: **Pilot Anopheles gambiae full-length cDNA study: sequencing and initial characterization of 35,575 clones.** *Genome Biol* 2005, **6(4):**R39.
15. Bonaldo MF, Lennon G, Soares MB: **Normalization and subtraction: two approaches to facilitate gene discovery.** *Genome Res* 1996, **6(9):**791-806.
16. Ewing B, Green P: **Analysis of expressed sequence tags indicates 35,000 human genes.** *Nat Genet* 2000, **25(2):**232-234.
17. Gilchrist MJ, Zorn AM, Voigt J, Smith JC, Papalopulu N, Amaya E: **Defining a large set of full-length clones from a Xenopus tropicalis EST project.** *Dev Biol* 2004, **271(2):**498-516.
18. **Wellcome X. tropicalis Full-Length Database** [http://informatics.gurdon.cam.ac.uk/online/xt-fl-db.html]
19. Klein SL, Strausberg RL, Wagner L, Pontius J, Clifton SW, Richardson P: **Genetic and genomic tools for Xenopus research: The NIH Xenopus initiative.** *Dev Dyn* 2002, **225(4):**384-391.
20. Ewing RM, Ben Kahla A, Poirot O, Lopez F, Audic S, Claverie JM: **Large-scale statistical analyses of rice ESTs reveal correlated patterns of gene expression.** *Genome Res* 1999, **9(10):**950-959.
21. Tata JR: **Amphibian metamorphosis as a model for the developmental actions of thyroid hormone.** *Mol Cell Endocrinol* 2006, **246(1-2):**10-20.
22. Romualdi C, Bortoluzzi S, Danieli GA: **Detecting differentially expressed genes in multiple tag sampling experiments: comparative evaluation of statistical tests.** *Hum Mol Genet* 2001, **10(19):**2133-2141.
23. Gibson BW, Poulter L, Williams DH, Maggio JE: **Novel peptide fragments originating from PGLa and the caerulein and xenopsin precursors from Xenopus laevis.** *J Biol Chem* 1986, **261(12):**5341-5349.
24. Izutsu Y, Tochinai S, Maeno M, Iwabuchi K, Onoe K: **Larval antigen molecules recognized by adult immune cells of inbred Xenopus laevis: partial characterization and implication in metamorphosis.** *Dev Growth Differ* 2002, **44(6):**477-488.
25. Seki T, Kikuyama S, Yanaihara N: **Development of Xenopus laevis skin glands producing 5-hydroxytryptamine and caerulein.** *Cell Tissue Res* 1989, **258(3):**483-489.
26. **Xtscope** [http://indigene.ibaic.u-psud.fr/EST]
27. Maeda N, Kasukawa T, Oyama R, Gough J, Frith M, Engstrom PG, Lenhard B, Aturaliya RN, Batalov S, Beisel KW, Bult CJ, Fletcher CF, Forrest AR, Furuno M, Hill D, Itoh M, Kanamori-Katayama M, Katayama S, Katoh M, Kawashima T, Quackenbush J, Ravasi T, Ring BZ, Shibata K, Sugiura K, Takenaka Y, Teasdale RD, Wells CA, Zhu Y, Kai C, Kawai J, Hume DA, Carninci P, Hayashizaki Y: **Transcript annotation in FANTOM3: mouse gene catalog based on physical cDNAs.** *PLoS Genet* 2006, **2(4):**e62.
28. Pollet N, Muncke N, Verbeek B, Li Y, Fenger U, Delius H, Niehrs C: **An atlas of differential gene expression during early Xenopus embryogenesis.** *Mech Dev* 2005, **122(3):**365-439.
29. Pollet N, Schmidt HA, Gawantka V, Vingron M, Niehrs C: **Axeldb: a Xenopus laevis database focusing on gene expression.** *Nucleic Acids Res* 2000, **28(1):**139-140.
30. Rungger D: **Xenopus helveticus, an endangered species?** *Int J Dev Biol* 2002, **46(1):**49-63.
31. Bulle F, Chiannilkulchai N, Pawlak A, Weissenbach J, Gyapay G, Guellaen G: **Identification and chromosomal localization of human genes containing CAG/CTG repeats expressed in testis and brain.** *Genome Res* 1997, **7(7):**705-715.
32. Eisen MB, Spellman PT, Brown PO, Botstein D: **Cluster analysis and display of genome-wide expression patterns.** *Proc Natl Acad Sci U S A* 1998, **95(25):**14863-14868.
33. **JGI X. tropicalis v4.1 home** [http://genome.jgi-psf.org/Xentr4/Xentr4.download.html]
34. **SNP/INDEL Discovery Pipeline based on CAP3 assembly** [http://cgpdb.ucdavis.edu/SNP_Discovery/]